\documentclass[aps,prl, twocolumn, superscriptaddress, longbibliography]{revtex4-2}

\usepackage{times}
\usepackage{amsmath}
\usepackage{amssymb}
\usepackage{graphicx}
\usepackage[caption=false, position=top, singlelinecheck=off, justification=raggedright]{subfig}
\usepackage{color}
\usepackage{pst-node}
\usepackage[unicode]{hyperref}
\hypersetup{unicode=true, colorlinks=true, linkcolor=blue, citecolor=blue}
\usepackage{float}
\usepackage[normalem]{ulem}

\begin{document}

\title{Torus bifurcation of a dissipative time crystal}

\author{Jayson G. Cosme}
\thanks{These authors have contributed equally to this work.}
\affiliation{National Institute of Physics, University of the Philippines, Diliman, Quezon City 1101, Philippines}

\author{Phatthamon Kongkhambut}
\thanks{These authors have contributed equally to this work.}
\affiliation{Center for Optical Quantum Technologies and Institute for Quantum Physics, Universit\"at Hamburg, 22761 Hamburg, Germany}
\affiliation{Quantum Simulation Research Laboratory, Department of Physics and Materials Science, Faculty of Science, Chiang Mai University, Chiang Mai, 50200, Thailand}
\affiliation{Thailand Center of Excellence in Physics, Office of the Permanent Secretary, Ministry of Higher Education, Science, Research and Innovation, Thailand}

\author{Anton B\"{o}lian}
\affiliation{Center for Optical Quantum Technologies and Institute for Quantum Physics, Universit\"at Hamburg, 22761 Hamburg, Germany}

\author{Richelle Jade L. Tuquero}
\affiliation{National Institute of Physics, University of the Philippines, Diliman, Quezon City 1101, Philippines}

\author{Jim Skulte}
\affiliation{Center for Optical Quantum Technologies and Institute for Quantum Physics, Universit\"at Hamburg, 22761 Hamburg, Germany}
\affiliation{The Hamburg Center for Ultrafast Imaging, Luruper Chaussee 149, 22761 Hamburg, Germany}

\author{Ludwig Mathey}
\affiliation{Center for Optical Quantum Technologies and Institute for Quantum Physics, Universit\"at Hamburg, 22761 Hamburg, Germany}
\affiliation{The Hamburg Center for Ultrafast Imaging, Luruper Chaussee 149, 22761 Hamburg, Germany}

\author{Andreas Hemmerich}
\email[]{andreas.hemmerich@uni-hamburg.de}
\affiliation{Center for Optical Quantum Technologies and Institute for Quantum Physics, Universit\"at Hamburg, 22761 Hamburg, Germany}
\affiliation{The Hamburg Center for Ultrafast Imaging, Luruper Chaussee 149, 22761 Hamburg, Germany}

\author{Hans Ke{\ss}ler}
\email[]{hans.g.kessler@gmail.com}
\affiliation{Center for Optical Quantum Technologies and Institute for Quantum Physics, Universit\"at Hamburg, 22761 Hamburg, Germany}
\affiliation{Physikalisches Institut, Rheinische Friedrich-Wilhelms-Universit\"at, 53115  Bonn, Germany}


\begin{abstract}   
{Using a quantum gas setup consisting of a Bose-Einstein condensate strongly coupled to a high-finesse optical cavity by a
transverse pump laser, we experimentally observe an instability of a dissipative continuous time crystal (CTC) towards a time crystalline state exhibiting two prominent oscillation frequencies. Applying a mean-field approximation model and a Floquet analysis, we theoretically confirm that this transition is a manifestation in a many-body system of a torus bifurcation between a limit cycle (LC) and a limit torus (LT).} We theoretically illustrate the LC and LT attractors using the minimal model and experimentally reconstruct them using Takens' embedding theorem applied to the non-destructively measured intracavity photon dynamics.
\end{abstract}

\maketitle 

\textit{Introduction.---} Critical transitions are ubiquitous in complex dynamical systems in nature. In condensed matter systems, a mean-field approximation of a many-body system can lead to a few-mode model, which maps certain aspects of a phase transition onto the mathematical framework of critical transitions and bifurcation theory commonly applied to classical nonlinear systems \cite{Strogatz2000}. For example, in dissipative systems, various phase transitions can be understood as critical transitions between two fixed points or a fixed point and a periodic solution \cite{Kirton2018, Iemini2018, Lledo2019, larson2021, DickeModel,Stitely2020, Seibold2020, Seibold2022, Buca2022, Kosior2023, Skulte2024, Alaeian2024}. A physical system known to host a particularly rich variety of phase transitions is the quantum-gas cavity-QED platform \cite{Ritsch2013,Mivehvar2021}. In the transversely pumped atom-cavity scenario, a pitchfork bifurcation from a trivial to a nontrivial fixed point is realised in the normal phase (NP) to superradiant phase (SP) transition as the strength of the light-matter interaction is increased \cite{Baumann2010, Klinder2015, Zhang2021, Helson2023}. The emergence of a continuous time crystal (CTC), i.e., another many-body state, in that same scenario \cite{Kongkhambut2022} can be understood as a Hopf bifurcation between the SP and a limit cycle (LC) as the light-matter interaction is further increased in a mean-field description of the system \cite{Piazza2015, Kessler2019, Kessler2020, Gao2023, Skulte2024}. Due to this correspondence between the many-body phase, the CTC, and its mean-field counterpart, the LC, we will refer to CTCs as LCs, hereafter. Within bifurcation theory, a LC can also become unstable as the system goes through a route to chaos \cite{Strogatz2000}. An example is the period-doubling instability of a LC observed in a Rydberg atomic gas \cite{Liu2024}. A less common type of instability of a LC, known to arise in classical systems \cite{Kuznetsov2004, Sacker2009, Sinha2006, Nicolaou2021, Grziwotz2023}, but yet to be experimentally demonstrated in quantum systems despite a theoretical proposal \cite{Yusipov2019}, is the Neimark-Sacker, secondary Hopf, or torus bifurcation. This bifurcation is characterised by a LC becoming unstable towards the formation of a limit torus (LT), which can be detected by observing the associated quasiperiodic dynamics. Quasiperiodic dynamics has been previously predicted and observed in periodically driven systems \cite{Autti2018, Giergiel2019, Pizzi2019, He2024}. 

In this work, we experimentally demonstrate that for strong light-matter interactions, the LC previously observed in an atom-cavity system \cite{Kongkhambut2022, Skulte2024}, comprising a Bose-Einstein condensate (BEC) inside an recoil-resolved optical resonator \cite{Kessler2014, Klinder2016}, becomes unstable, leading to a quasiperiodic dynamical state identified as a LT. The emergence of a LT corresponds to the appearance of a second spectral component in the Fourier spectrum of the intracavity photon dynamics, which is incommensurate with the response frequency of the LC. Supplementing our experimental results with numerical simulations, we infer from the behaviour of the Floquet multipliers of a minimal model of the system that the transition between a LC and LT can be understood as a Neimark-Sacker bifurcation, which constitutes its first observation in a quantum-coherent light-matter system.

\begin{figure*}[!ht]
\centering
\includegraphics[width=2\columnwidth]{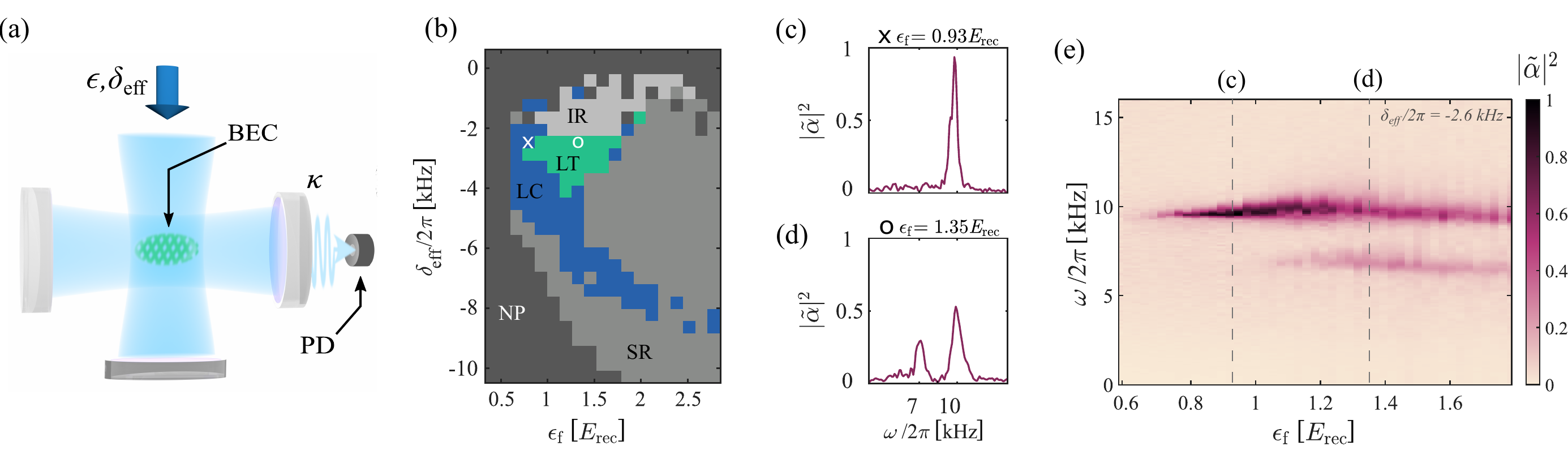}
\caption{(a) Schematic of the setup of a BEC transversally pumped with a standing wave potential. (b) Experimental phase diagram spanned by $\epsilon_f$ and $\delta_\mathrm{eff}$ showing distinct regimes: the (NP) normal phase, (SP) superradiant phase, (LC) limit cycle, (LT) limit torus, and (IR) irregular response. More details on how the phase diagram is constructed can be found in the supplementary material \cite{suppmat}. (c), (d) Experimentally measured Fourier spectra of the intracavity photon number for (c) a  LC and (d) a LT. (e) Fourier spectra for varying pump intensities at fixed detuning $\delta_\mathrm{eff} = - 2\pi \times 2.6~\mathrm{kHz}$. The vertical lines denote the pump intensities corresponding to (c) and (d).} 
\label{fig:1} 
\end{figure*}

\textit{Experiment.---} In the experiment, we prepare a BEC of $^{87}$Rb atoms with {a} particle number of around $N_\mathrm{a}\approx 4\times10^4$. The BEC is strongly coupled to the fundamental mode of an optical high-finesse cavity by pumping with a retro-reflected light field along the direction perpendicular to the cavity axis as depicted in Fig.~\ref{fig:1}(a). We pump with a wavelength $\lambda_\mathrm{p} = 791.59\,$nm, which is blue-detuned with respect to the $D_1$-line of $^{87}$Rb at $794.98\,$nm, and thus the pump field provides a repulsive standing wave potential for the atoms. The recoil frequency corresponding to this pump wavelength is $\omega_\mathrm{rec} = 2\pi \times 3.7~\mathrm{kHz}$. The lattice depth produced by a single photon scattered into the cavity is $U_\mathrm{0} = 2\pi\times0.7\,$Hz. The bare detuning between the pump and cavity frequencies is $\delta_c$. When accounting for the the collective light shift, this gives rise to an effective detuning $\delta_\mathrm{eff} = \delta_c - N_\mathrm{a} U_0/2$. The cavity field decay rate is $\kappa=2\pi\times 3.2$~kHz. Thus, the system operates in the recoil-resolved regime $\kappa \sim \omega_\mathrm{rec}$, which means that the photon-mediated infinite-range interaction between the atoms is dynamical, and the cavity dynamics cannot be adiabatically eliminated, as for broad-bandwidth cavities with $\kappa \gg \omega_\mathrm{rec}$ \cite{Mivehvar2021}.

In Fig.~\ref{fig:1}(b), the experimental phase diagram in the parameter space spanned by the final pump intensity $\epsilon_\mathrm{f}$ and the effective detuning $\delta_{\mathrm{eff}}$ is shown. We note that the sharp boundaries between the phases are due to the set of criteria used to distinguish between the normal phase (NP), superradiant phase (SP), and irregular response (IR) as discussed in the supplemental material \cite{suppmat}. In the following, we will focus on the LC and LT. Later, we will demonstrate theoretically that the LT in this work appears as a torus in three-dimensional coordinate space. First, we will discuss the experimental signatures of the LC and LT based on the dynamics of the intracavity photon number $|a|^2$ used to construct the phase diagram in Fig.~\ref{fig:1}(b). 

We analyse the normalised Fourier spectra of $|a|^2$, {with some examples shown in} Figs.~\ref{fig:1}(c) and \ref{fig:1}(d) for a LC and LT, respectively. Both the LC and LT have well-defined oscillation frequencies appearing as peaks in the Fourier spectra of $|a|^2$. In a LC, there is one dominant response frequency, in addition to possible higher harmonics, as seen from the peak around $\sim$10 kHz in Fig.~\ref{fig:1}(c). In contrast, in a LT, a second prominent frequency peak is found, as shown in the subdominant peak around $\sim$7 kHz in Fig.~\ref{fig:1}(d). We emphasise that an incommensurate ratio between the primary and secondary peaks is indicative of a quasiperiodic dynamics, a defining feature of a LT. Note that a period-doubling bifurcation occurs when the ratio between the secondary and primary peaks is exactly $1/2$. In the Supplemental Material \cite{suppmat}, we explain in detail how we experimentally classify LC and LT based on their unique signatures.

For fixed $\delta_\mathrm{eff} = - 2\pi \times 2.6~\mathrm{kHz}$ and increasing pump intensity, we show in Fig.~\ref{fig:1}(e) that a LC becomes unstable and turns into a LT as a second peak in the spectra emerges. This is also supported by the island of LT found for higher values of $\epsilon_\mathrm{f}$ in the phase diagram Fig.~\ref{fig:1}(b). Therefore, we experimentally observe an instability of a LC towards a quasiperiodic dynamical state for increasingly strong light-matter interactions between the cavity photons and the atoms, the strength of which is effectively controlled by the pump intensity.

\begin{figure*}[!ht]
\centering
\includegraphics[width=1.8\columnwidth]{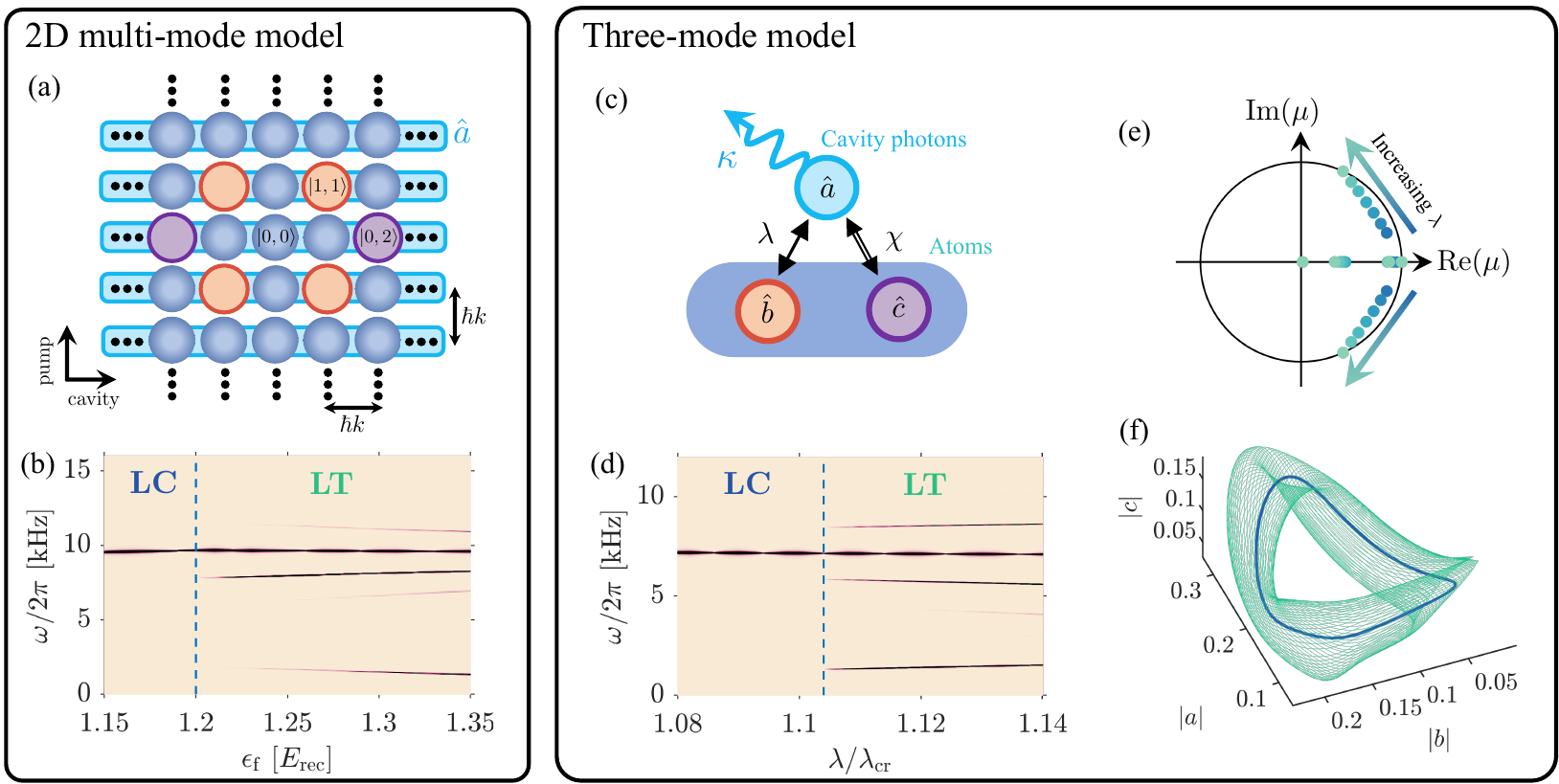}
\caption{{Mean-field theory results. In (a), (b), a 2D model with 169 momentum modes and a single cavity mode is applied. The relevant modes associated with the momentum excitations in the two-mode approximation of the atomic sector for the three-mode model in (c) follow the same color scheme. In (c)-(f) a minimal model with few coupled bosonic modes is used.} (a), (c) Sketch of the theoretical models. (b), (d) Theoretical Fourier spectra for (b) $\delta_\mathrm{eff} = - 2\pi \times 5~\mathrm{kHz}$ against final pump intensity $\epsilon_\mathrm{f}$ and (d)  $-\delta_\mathrm{eff}=\omega_p=\kappa$. The maximum of the Fourier spectra is capped at $10^{-2}$ to emphasize the appearance of well-defined side peaks that are incommensurate with the dominant oscillation frequency of the LC. (e) Floquet multipliers $\mu$ for increasing $\lambda/\lambda_\mathrm{SP} \in [1.05,1.10]$ according to the minimal model. (f) Dynamics in a space spanned by $|a|$, $|b|$, and $|c|$ for $\lambda$ slightly smaller (dark) and larger (light) than the critical value for a LT to emerge.} 
\label{fig:2} 
\end{figure*}

\textit{Multi-mode simulation.---} To shed light on the nature of the transition between the LC and LT, we turn to theoretical descriptions of the transversely pumped atom-cavity system. {We first employ a mean-field approach in a multimode expansion in momentum space. Later, we compare this approach to a few-mode mean-field approximation, which we discuss within a nonlinear dynamics methodology. For the multimode approach, we use a two-dimensional (2D) simplified model, considering only atomic momentum excitations along the pump and cavity directions together with the single cavity mode, as depicted in Fig.~\ref{fig:2}(a). We expand the atomic field operators using a plane-wave basis as done in Refs.~\cite{Cosme2019,Kongkhambut2024}.}

In our numerical simulations, we use the set of experimental parameters described earlier for an initial homogeneous BEC and {a} slightly occupied cavity mode $|a|^2(t=0) = 10^{-3}$ to ensure that the system can be pushed out of the NP. We present in Fig.~\ref{fig:2}(b) the Fourier spectra for different pump intensities and fixed detuning $\delta_\mathrm{eff} = - 2\pi \times 5~\mathrm{kHz}$ taken at $t \in [50,100]~\mathrm{ms}$; see Supplemental Material for the corresponding numerical phase diagram \cite{suppmat}. Consistent with the experimental results in Fig.~\ref{fig:1}(e), we also observe the appearance of side peaks that are incommensurate with the oscillation frequency of the previously stable LC before it turns into a LT for stronger $\epsilon$. Thus, our 2D model correctly predicts the instability observed in the experiment. {We compare the predictions of the multimode model and the experimental observations to a few-mode model.  In~\cite{Skulte2024} we have demonstrated that it captures the Hopf bifurcation from a SP to a LC. As such, we will next focus on this minimal model to show that the LC-LT transition is a torus bifurcation and to show that a three-dimensional coordinate space is enough to reconstruct the LC and LT attractors.}

\textit{Three-mode model.---} The Hamiltonian for the minimal bosonic model reads \cite{Skulte2024}
\begin{align}
\label{eq:H}
\hat{H} &= \omega_p \hat{a}^\dagger \hat{a} + \omega_{10} \hat{b}^\dagger \hat{b}+\omega_{20} \hat{c}^\dagger \hat{c} +\lambda \left(\hat{a}^\dagger+\hat{a} \right)\left(\hat{b}^\dagger+\hat{b} \right) \\ \nonumber
&\qquad + \chi \hat{a}^\dagger \hat{a} \left(\hat{c}^\dagger+\hat{c} \right).
\end{align}
In the context of the atom-cavity system, the operator $\hat{a}^\dagger$ creates a photon in the cavity mode. Excitations from the ground- state BEC mode $|0,0\rangle$ to the momentum modes  $|\pm \hbar k,\pm \hbar k\rangle$, crucial to the formation of a density wave pattern in the SP,  are implemented via the bosonic creation operator $\hat{b}^\dagger$. The dynamical optical lattice formed along the cavity direction leads to $2\hbar k$-momentum excitations from $|0,0\rangle$ to $|0,\pm 2 \hbar k\rangle$, which are represented by the creation operator $\hat{c}^\dagger$. 
This excitation channel couples to the intracavity intensity, and the coupling strength is controlled by the single photon lattice depth $U_0$ included in the prefactor $\chi$ in Eq.~\eqref{eq:H} \cite{Skulte2024}. We point out that this term is unique to our minimal model Eq.~\eqref{eq:H} and is absent in the usual treatment of a NP in the standard Dicke model in the thermodynamic limit \cite{Jara2024}. The remaining parameters in Eq.~\eqref{eq:H} are the photon detuning $\omega_p = -\delta_\mathrm{eff}$, the standard light-matter coupling strength $\lambda$ proportional to the amplitude of the pump field, and the bare transition frequencies $\omega_{10}$ and $\omega_{20}$ corresponding to transitions from the ground state $|0,0\rangle$ to $|\pm \hbar k,\pm \hbar k\rangle$  and $|0,\pm 2 \hbar k\rangle$, respectively. A mean-field approximation of this model leads to a set of equations of motion (EOM) for $a=\langle \hat{a} \rangle$, $b=\langle \hat{b} \rangle$, and $c=\langle \hat{c} \rangle$ given by
\begin{align}\label{eq:eom3mode}
\frac{\mathrm{d} a}{\mathrm{d} t} &= -i \left( \omega_p-i\kappa+\chi \left(c+c^* \right) \right) a-i\lambda \left(b+b^* \right),  \\ \nonumber
\frac{\mathrm{d} b}{\mathrm{d} t} &= -i \omega_{10} b -i \lambda  \left(a+a^* \right), \\ \nonumber
 \frac{\mathrm{d} c}{\mathrm{d} t} &= -i \omega_{20} c -i \chi  a^* a~.
\end{align}

We numerically propagate the above EOM for an initial configuration with slight occupation of the cavity mode and excited states, $a=b=c=10^{-3}$. The parameters are consistent with the experiment, i.e., $\omega_{10}=2\omega_\mathrm{rec} = \omega_{20}/2$, $\omega_p = -\delta_\mathrm{eff}$, and $\chi \approx 4\kappa$ \cite{Skulte2024}. We explore the dynamics for varying light-matter coupling strength in units of the critical value for the NP-SP transition, $\lambda_\mathrm{cr} = \sqrt{(\kappa^2+\omega_p^2)\,\omega_{10}/4\omega_{p}}$. In Fig.~\ref{fig:2}(d), we present the Fourier spectra of $|a|^2$ for the minimal model taken at {$\omega_{10} \, t \in [1,2] \times10^3$.} We point out the similarity in the general structure of the Fourier spectra of the 2D atom-cavity and minimal models, Figs.~\ref{fig:2}(b) and \ref{fig:2}(d), which corroborates the applicability of our minimal model for gaining further insights about the LC-LT transition. We note that while the three-mode model in Eq.~\eqref{eq:eom3mode} can be further simplified to a model with two modes and a Kerr nonlinearity by an adiabatic elimination of the dynamics of the $c$-mode, such a model fails to capture the emergence of a Neimark-Sacker bifurcation and, instead, predicts a period-doubling bifurcation \cite{suppmat}. This highlights the crucial role played by the third atomic level for the existence of a Neimark-Sacker bifurcation in the system.

\textit{Floquet analysis.---} Unlike in Ref.~\cite{Skulte2024}, which focused on the instability of the SP, a fixed point in the language of dynamical systems \cite{Strogatz2000}, we aim to identify in this work the type of bifurcation that leads to the instability of a LC, a periodic solution. As such, we employ a Floquet stability analysis \cite{Patra2019a, Patra2019b, Kuznetsov2004, Fruchart2021} of the LCs, which we briefly summarise as follows. First, we write Eq.~\eqref{eq:eom3mode} as $\partial_t \mathbf{X} = \mathbf{F}(\mathbf{X})$, where $\mathbf{X} = [a,a^*,b,b^*,c,c^*]^{\top}$, and numerically obtain the LC solution $\mathbf{X}_\mathrm{LC}(t)$ with period $T$. Next, we linearise Eq.~\eqref{eq:eom3mode} with respect to small perturbations around the stable LC solution, $\mathbf{X}_\mathrm{LC}(t) + \delta_\mathbf{X}$, which yields a linearised EOM $\partial_t \delta_\mathbf{X} = \mathbf{J}_{\mathrm{LC}}(t) \delta_\mathbf{X}$. Here, we have introduced the time-periodic Jacobian matrix $\mathbf{J}_{\mathrm{LC}}(t)$, which is the standard Jacobian matrix evaluated at the LC solution $\mathbf{J}_{\mathrm{LC}}(t) = \mathbf{J}_0\lvert_{\mathbf{X}_\mathrm{LC}}$, see Supplemental Material for an expression of the Jacobian matrix $\mathbf{J}_0$. We numerically obtain the solution of the linearised equation $\delta^{(1)}_\mathbf{X}(t)$ for an initial state that samples the vacuum fluctuations of the $a$-, $b$-, and $c$-modes, e.g., $a \in (\xi_1 + i \xi_2)/2\sqrt{N_\mathrm{a}}$, where $\xi_i$ is a random number drawn from a standard Gaussian distribution. We gather the solution as a column vector $[\delta^{(1)}_\mathbf{X}(t)]$ with $N=6$ rows corresponding to the dimension of the matrix $\mathbf{J}_0$ and the elements of which are the solutions $\delta^{(1)}$ evaluated at a specific time $t$. We obtain $N$ linearly independent solutions for $\delta_\mathbf{X}$ using a set of $N$ randomly picked initial states to construct the so-called fundamental matrix $\mathbf{\Delta}(t) = \biggl[ \left[\delta^{(1)}_\mathbf{X}(t)\right], \dots, \left[\delta^{(N)}_\mathbf{X}(t)\right]\biggr]$. Finally, we construct the monodromy matrix $\mathbf{M} = \left( \mathbf{\Delta}(t)\right)^{-1} \mathbf{\Delta}(t+T)$, the eigenvalues of which are the Floquet multipliers $\mu_i$ with the property $\delta_\mathbf{X}(t+T) = \mu_j \delta_\mathbf{X}(t)$. Note that $\mu_j$ is, in general, a complex number as the monodromy matrix is typically non-Hermitian. For periodic solutions such as a LC, one Floquet multiplier $\mu_0$ is marginal or trivial as it is always equal to unity, $\mu_0 = 1$ \cite{Patra2019a,Patra2019b,Fruchart2021}. This is because $\delta_\mathbf{X} = \dot{\mathbf{X}}_\mathrm{LC}$ is also a periodic solution of the linearised EOM and it has the same period $T$ as the LC. Excluding the marginal multiplier, a LC is stable if $|\mu_j|<1~\forall~j \neq 0$ \cite{Kuznetsov2004}. On the other hand, if at least one of the Floquet multipliers has $|\mu_j| \geq 1$, then the LC is unstable. 

Instability of a periodic solution is deduced if at least one nontrivial $\mu_j$ crosses the unit circle when the multipliers are plotted in a complex plane. For example, a period-doubling instability of a LC, as in Ref.~\cite{Liu2024}, occurs when a purely real Floquet multiplier $\mu_j \to -1$ as the parameter for the transition is tuned \cite{Kuznetsov2004, Grziwotz2023}. In fact, our simulations for a 1D atom-cavity system \cite{suppmat} predict a period-doubling instability of the LC, which suggests that an appropriate choice of transition frequencies in other setups \cite{Liu2024, Greilich2024} could lead to realisation of other types of bifurcations. Here, for transition frequencies consistent with the appropriate excitation frequencies of the full atom-cavity setup, the behaviour of the Floquet multipliers for increasing light-matter coupling strength $\lambda$ is consistent with a Neimark-Sacker or torus bifurcation \cite{Kuznetsov2004, Grziwotz2023}. This is evident from Fig.~\ref{fig:2}(e), in which we show the Floquet multipliers for different values of $\lambda$. Indeed, as $\lambda$ increases, a pair of $\mu_j$ eventually crosses the unit circle at an angle meaning $\mathrm{Im}(\mu_j)>0$, consistent with a torus bifurcation \cite{Kuznetsov2004,Grziwotz2023}. 

\begin{figure}[!htpb]
\centering
\includegraphics[width=\columnwidth]{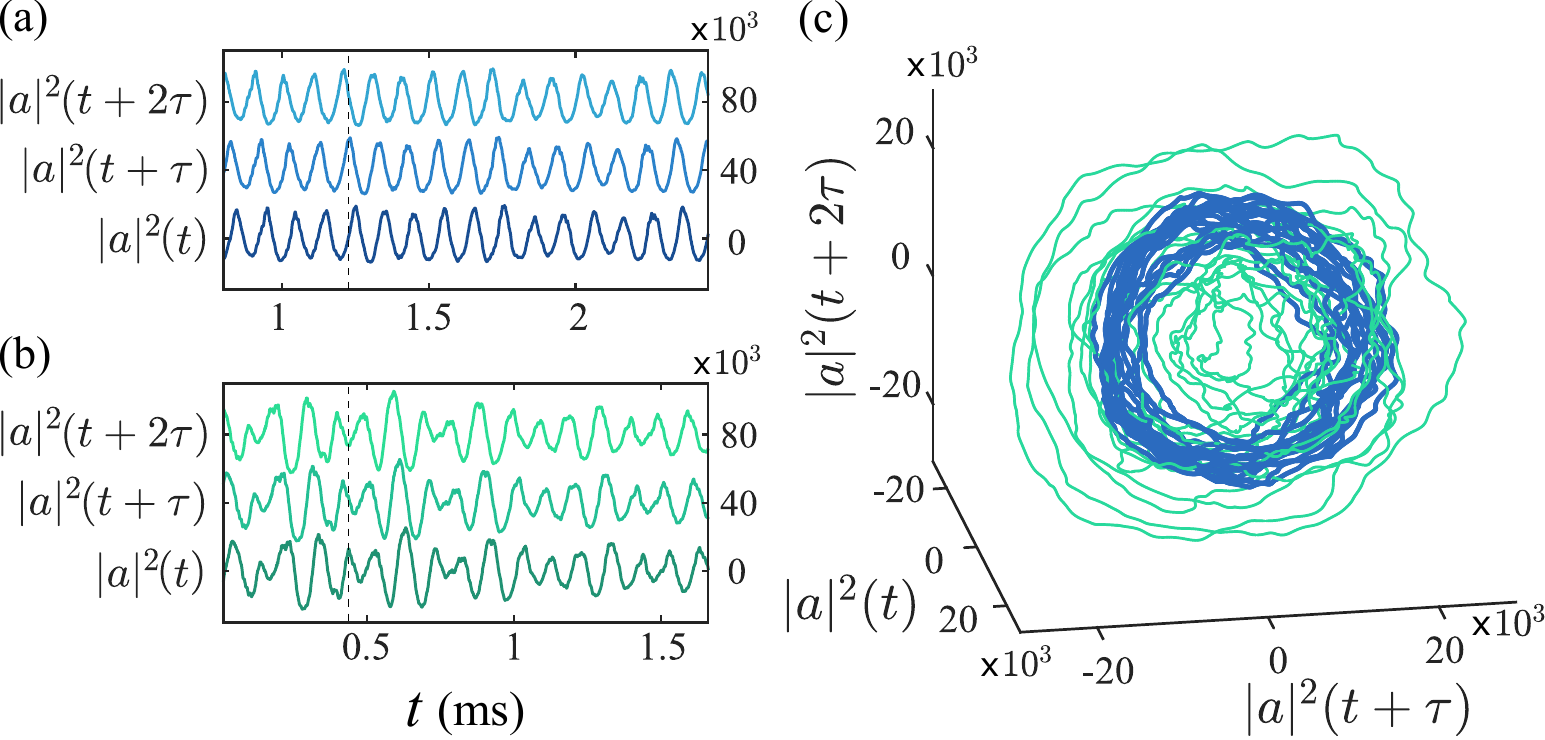}
\caption{{Singlet trajectory measurement of the dynamics of the cavity photon occupation $|a|^2$ for (a) a LC and (b) a LT. The parameters for (a) and (b) are the same as those marked as 'X' and 'O' in Fig.~\ref{fig:1}(b). Each trajectory corresponds to the same data, albeit delayed by $\tau$ and $2\tau$, and shifted vertically by $40 \times 10^3$ and $80 \times 10^3$ from the original signal which is set to oscillate around zero, respectively. The time delay is $\tau = 0.2$ ms. Vertical lines in (a) and (b) are guide to the eye to emphasize the time delay. (c) Experimental reconstruction of the LC and LT attractors using Takens' embedding theorem using the time-delayed signals in (a) and (b).}}
\label{fig:3} 
\end{figure}

\textit{Attractor reconstruction.---} The attractor of a stable LC forms a closed loop in coordinate space while a LT forms a torus. According to our minimal model, a three-dimensional coordinate space spanned by $|a|$, $|b|$, and $|c|$ is sufficient to illustrate the LC and LT attractors in the atom-cavity system, as demonstrated in Fig.~\ref{fig:2}(f). This is especially insightful for the experimental reconstruction of the attractors in the atom-cavity setup, wherein the large number of atomic modes relevant in the dynamics makes it difficult to identify the minimum dimension of the coordinate space to show the attractor. Moreover, another limiting factor is that the destructive measurement of the dynamics of the atomic modes leads to a coarse scan of their time evolution. Nevertheless, attractors can be reconstructed using time-delayed or lagged coordinates \cite{Takens,Grziwotz2023,Greilich2024}, which we can exploit together with the \textit{in situ} monitoring of the cavity photon dynamics. Following Takens' theorem, we delayed the single-shot intracavity photon occupation dynamics $|a|^2$ by $\tau=0.02$ ms and $2\tau$, as depicted in Figs.~\ref{fig:3}(a) and \ref{fig:3}(b), and create a three-dimensional lagged coordinates for reconstructing the LC and LT attractors. The experimentally reconstructed attractors are shown in Fig.~\ref{fig:3}(c), wherein the trajectory for a LT is found to occupy a larger volume in the lagged-coordinate system than the LC. {We attribute the apparent spiralling of the trajectory towards the centre of the LT attractor in Fig.~\ref{fig:3}(c)} to atom losses, even though, in an idealized scenario the attractor should just remain a torus in the coordinate space, as highlighted by the theoretical results in Fig.~\ref{fig:2}(f).

\textit{Conclusion.---} In conclusion, we have experimentally demonstrated a transition between two distinct time crystalline phases in a quantum gas setup. When approximated at the mean-field level, this transition exemplifies a Neimark-Sacker, or torus bifurcation, first predicted in nonlinear systems but now observed in a quantum many-body system. The torus bifurcation is observed when the CTC, initially defined by a single oscillation frequency, becomes unstable and begins to exhibit quasiperiodic oscillations characterised by two dominant incommensurate frequencies. Theoretical analyses based on a minimal model and Floquet theory confirm that the quasiperiodic dynamical state is a LT and that the LC-LT transition can be understood as a torus bifurcation in the context of nonlinear dynamical systems. Our work opens up the possibility to utilize the atom-cavity platform considered here or other similar driven-dissipative quantum systems operating in the few-body or mesoscopic regime \cite{Ho2024} as a natural testbed for exploring the quantum nature of bifurcations, limit cycles, and synchronization \cite{Arosh2021, Waechtler2023, Dutta2024}. On the conceptual level, regarding phase transitions in and out of time crystalline states, we have provided support for the view that the criticality of time crystalline states is captured by bifurcation theory at the mean-field level. Finally, our work provides an elementary platform for observing quasiperiodic dynamical phases, a topic of growing interest in the study of time crystal scenarios \cite{He2024, Huang2024}, wherein dissipation plays a pivotal role.

\begin{acknowledgments}
We thank J. Klinder and C. Georges for their support during the early stage of the project. P.K. would like to thank P. Simakachorn for useful discussions. This work was supported by the QuantERA II Programme that has received funding from the European Union's Horizon 2020 research and innovation programme under Grant Agreement No. 101017733, the Deutsche Forschungsgemeinschaft (DFG, German Research Foundation) ``SFB-925" project 170620586, and the Cluster of Excellence ``Advanced Imaging of Matter" (EXC 2056), Project No. 390715994. The project is co-financed by ERDF of the European Union and by `Fonds of the Hamburg Ministry of Science, Research, Equalities and Districts (BWFGB)'. J.G.C. and R.J.L.T. acknowledge support from the DOST-ASTI's COARE high-performance computing facility. H.K. acknowledges funding by the state of North Rhine-Westphalia through the EIN Quantum NRW program. J.S. acknowledges support from the German Academic Scholarship Foundation.
\end{acknowledgments}

\nocite{Tuquero2024,Pizzi2021,Kessler2021}

\bibliography{references}

\end{document}


\title{Supplemental Material for \\  Torus bifurcation of a dissipative time crystal}

\author{Jayson G. Cosme}
\thanks{These authors have contributed equally to this work.}
\affiliation{National Institute of Physics, University of the Philippines, Diliman, Quezon City 1101, Philippines}

\author{Phatthamon Kongkhambut}
\thanks{These authors have contributed equally to this work.}
\affiliation{Center for Optical Quantum Technologies and Institute for Quantum Physics, Universit\"at Hamburg, 22761 Hamburg, Germany}
\affiliation{Quantum Simulation Research Laboratory, Department of Physics and Materials Science, Faculty of Science, Chiang Mai University, Chiang Mai, 50200, Thailand}
\affiliation{Thailand Center of Excellence in Physics, Office of the Permanent Secretary, Ministry of Higher Education, Science, Research and Innovation, Thailand}

\author{Anton B\"{o}lian}
\affiliation{Center for Optical Quantum Technologies and Institute for Quantum Physics, Universit\"at Hamburg, 22761 Hamburg, Germany}

\author{Richelle Jade L. Tuquero}
\affiliation{National Institute of Physics, University of the Philippines, Diliman, Quezon City 1101, Philippines}

\author{Jim Skulte}
\affiliation{Center for Optical Quantum Technologies and Institute for Quantum Physics, Universit\"at Hamburg, 22761 Hamburg, Germany}
\affiliation{The Hamburg Center for Ultrafast Imaging, Luruper Chaussee 149, 22761 Hamburg, Germany}

\author{Ludwig Mathey}
\affiliation{Center for Optical Quantum Technologies and Institute for Quantum Physics, Universit\"at Hamburg, 22761 Hamburg, Germany}
\affiliation{The Hamburg Center for Ultrafast Imaging, Luruper Chaussee 149, 22761 Hamburg, Germany}

\author{Andreas Hemmerich}
\email[]{hemmerich@physnet.uni-hamburg.de}
\affiliation{Center for Optical Quantum Technologies and Institute for Quantum Physics, Universit\"at Hamburg, 22761 Hamburg, Germany}
\affiliation{The Hamburg Center for Ultrafast Imaging, Luruper Chaussee 149, 22761 Hamburg, Germany}

\author{Hans Ke{\ss}ler}
\email[]{hans.g.kessler@gamil.com}
\affiliation{Center for Optical Quantum Technologies and Institute for Quantum Physics, Universit\"at Hamburg, 22761 Hamburg, Germany}
\affiliation{Physikalisches Institut, Rheinische Friedrich-Wilhelms-Universit\"at , 53115  Bonn, Germany}

\date{\today}

\maketitle


\section{Experimental details}
The experimental setup, as sketched in Fig.~1(a) in the main text, is comprised of a magnetically trapped Bose-Einstein condensate of $N_\mathrm{a} = 4\times 10^4$  $^{87}$Rb atoms, dispersively coupled to a narrowband high-finesse optical cavity. The trap creates a harmonic potential with trap frequencies $\omega = 2\pi \times (83.0,72.2,23.7)$~Hz. The corresponding Thomas-Fermi radii of the ensemble are $(r_x,r_y,r_z) = (4.4, 5.1, 15.5)~\mu m$. These radii are significantly smaller than the size of the Gaussian-shaped pump beam, which has a waist of $w_\mathrm{pump}~\approx~125~\mu m$. The pump beam is oriented transversally with respect to the cavity axis and retro-reflected to form a standing wave potential. The cavity field has a decay rate of $\kappa \approx ~2\pi \times3.2\,$kHz, which is comparable to the recoil frequency $\omega_\mathrm{rec} = E_\mathrm{rec}/\hbar  =~2\pi \times3.7\,$kHz for a pump wavelength of $\lambda_\mathrm{p} = 791.54\,$nm. The pump laser is blue detuned with respect to the relevant atomic transition of $^{87}$Rb at 794.98~nm. The maximum light shift per atom is $U_0 = 2\pi \times 0.7~\mathrm{Hz}$.

\section{Construction of the experimental phase diagram}
\begin{figure*}[!ht]
	\centering
	\includegraphics[width=0.5\columnwidth]{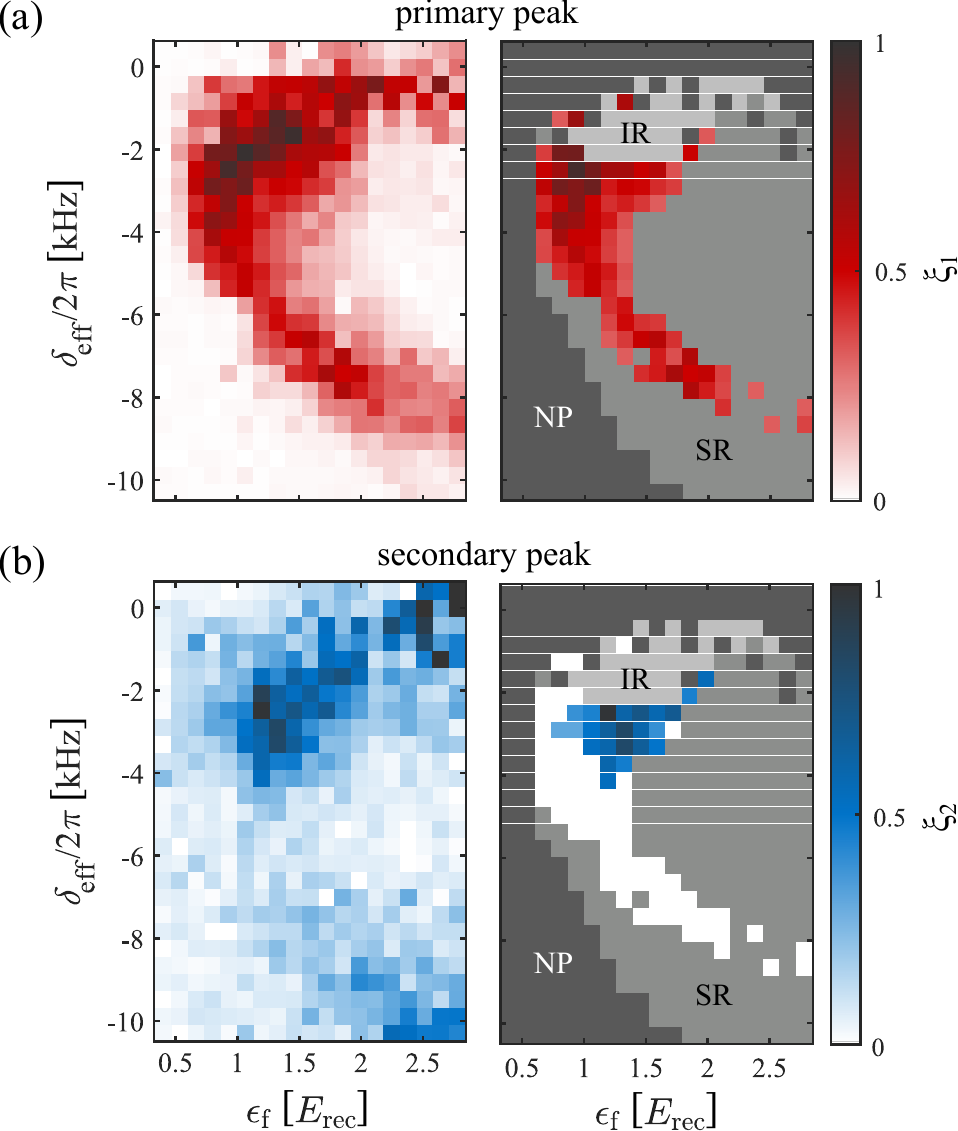}
	\caption{Primary and secondary peak amplitudes in the normalized Fourier spectra $|\tilde{\alpha}|^2$ of the detected intracavity photon number. Left: bare value. Right: with grey shading of each phase plotted on top: normal phase (NP), superradiant phase (SR), and irregular resonse regime (IR). (a) $\xi_{1}$ primary peak amplitude within the range of $[8,12]$~kHz. (b) $\xi_{2}$ secondary peak amplitude within the range of $[5,8]$~kHz.} 
	\label{sfig:1 phaseDiagram} 
\end{figure*}

To construct the experimental phase diagram shown in Fig.~1(b) in the main text, we used the following protocol. First, we fixed the effective pump-cavity detuning $\delta_{\mathrm{eff}}$, while linearly ramping the pump intensity to the final desired value $\epsilon_\mathrm{f}$ by a fixed ramp speed of $3.5 E_\mathrm{rec}/20 \mathrm{ms}$. Second, the final pump intensity $\epsilon_\mathrm{f}$ is held constant for 10 ms. We obtain the response of the system from the corresponding intracavity photon number $|\alpha|^2$ by detecting the light leaking out of one of the cavity mirror using a balanced heterodyne detector (see the next section for more details about the detection system). Then, we obtain the normalised single sided Fourier spectra $|\tilde{\alpha}|^2$ of the intracavity photon number during the 10 ms long hold time and obtain a primary and secondary peak amplitude, $\xi_\mathrm{1}$ and $\xi_\mathrm{2}$, which was called \textit{crystalline fraction} in the ref. \cite{Kessler2021}. $\xi_{1,2}$ are defined as the peak amplitudes of the Gaussian fits of the Fourier spectra within the range of $[8,12]$~kHz (LC's frequency range \cite{Kongkhambut2022}) and $[5,8]$~kHz, respectively. Left plots of Fig.~\ref{sfig:1 phaseDiagram}(a,b) show the obtained values of $\xi_\mathrm{1}$ and $\xi_\mathrm{2}$. Note that the region where both peaks coexist are potential candidates for limit torus phase (LT).

\begin{figure*}[!ht]
	\centering
	\includegraphics[width=0.45\columnwidth]{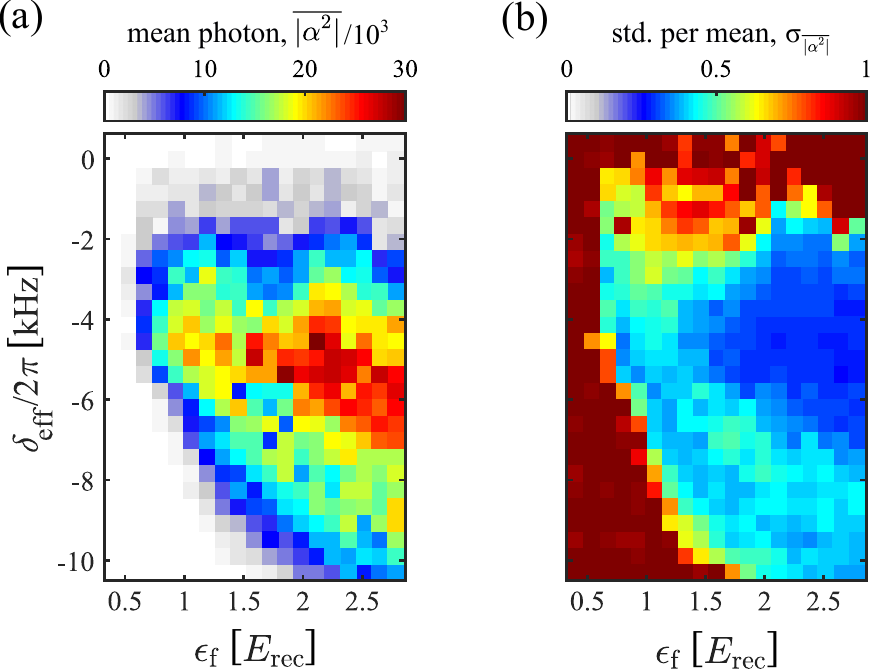}
	\caption{(a) Detected mean intracavity photon number $\overline{|\alpha|^2}$ during the first millisecond during the hold time. (b) Standard deviation of the detected intracavity photon number per mean $\sigma_{\overline{|\alpha|^2}} = \sigma_{|\alpha|^2}/\overline{|\alpha|^2}$ during the first millisecond of the hold time.} 
	\label{sfig:2 meanPhotons} 
\end{figure*}

The other important parameters obtained are the mean intracavity photon number $\overline{|\alpha|^2}$ and its standard deviation per mean $\sigma_{\overline{|\alpha|^2}} = \sigma_{|\alpha|^2}/\overline{|\alpha|^2}$, plotted in Fig.~\ref{sfig:2 meanPhotons}. They are obtained from $|\alpha|^2$ during the first milliseconds of the hold time. $\overline{|\alpha|^2}$ is used to distinguish the normal phase (NP), while $\sigma_{\overline{|\alpha|^2}}$ can tell how much the intracavity photon number fluctuates. 

To distinguish between the normal phase (NP), superradiant phase (SR), irregular response regime (IR), limit cycle (LC), and limit torus (LT), we take the following four quantities into account: $\overline{|\alpha|^2}$,  $\sigma_{\overline{|a|^2}}$, $\xi_\mathrm{1}$, and $\xi_\mathrm{2}$. Trajectories with a detected mean intracavity photon $\overline{|\alpha|^2}$ less than $2\times10^3$ photons are classified to be in the normal phase (NP). Note that $2\times10^3$ photons are the mean noise floor of the detection. We identify a superradiant phase (SR) if a trajectories has $\overline{|a|^2}>2\times10^3$ without any prominent primary and secondary peaks, which is quantified by $\xi_{1}$ and $\xi_{2}$ being below $1/e$ of the maximally detected values $\xi_{1,max}$ and $\xi_{2,max}$. The trajectories with $\overline{|a|^2}>2\times10^3$ but with high photon number fluctuations $\sigma_{\overline{|a|^2}}$ are classified as irregular response regime (IR), see the high fluctuation region in Fig.~\ref{sfig:2 meanPhotons}(b). We mark those regions that follow the above criteria as grey shaded areas in the right panels of Fig.~\ref{sfig:1 phaseDiagram}. 

LCs (blue area in Fig.~1(b) in the main text) are classified as the trajectories with $\overline{|a|^2}>2\times10^3$, low photon number fluctuations, and a prominent peak with crystalline fractions $\xi_{1}$ exceeding $1/e$ of the maximally detected values $\xi_{1,max}$. These trajectories are shown as the red area in Fig.~\ref{sfig:1 phaseDiagram}(a). Lastly, LTs (green area in Fig.~1(b) of the main text) are those trajectories satisfying the criteria of LCs but in addition, their secondary peak amplitude $\xi_{2}$ exceeds $1/e$ of the maximally detected value $\xi_{2,max}$. Their location in the phase diagram can be obtained by overlapping the remaining red and blue regions in the right panels of Figs.~\ref{sfig:1 phaseDiagram}(a) and \ref{sfig:1 phaseDiagram}(b).

\section{Cavity field detection}
Our experimental system is equipped with two detection setups for the light leaking out of the cavity. On one side of the cavity, we use a single photon counting module (SPCM), which provides access to the intensity of the intracavity field and the associated photon statistics. On the other side of the cavity, a balanced heterodyne detection setup is installed, which uses the pump beam as a local reference. The beating signal of the local oscillator with the light leaking out of the cavity allows for the observation of the time evolution of the intracavity photon number $|a|^2$ and the phase difference between the pump and the cavity field.


\section{Linearised dynamical equation} 
Linearising the equations of motion around the limit cycle (LC) mean-field solution $\mathbf{X}_\mathrm{LC} = [a(t),a^*(t),b(t),b^*(t),c(t),c^*(t)]^{\top}$, we obtain the linearised set of equations 
\begin{equation}
\label{eq:J}
\partial_t \delta_\mathbf{X} = \mathbf{J}_{\mathrm{LC}}(t) \delta_\mathbf{X}
\end{equation}
where $\mathbf{J}_{\mathrm{LC}}(t) = \mathbf{J}_0\lvert_{\mathbf{X}_\mathrm{LC}}$. Here, the Jacobian matrix is
\begin{widetext}
\begin{align}\label{eq:j0}
\mathbf{J}_0 = \frac{\partial \mathbf{F}(\mathbf{X})}{\partial \mathbf{X}} = \begin{pmatrix}
-i \left[\omega_p+\chi \left( c+c^* \right) \right]-\kappa & 0 & -i \lambda & -i\lambda & -i\chi a & -i\chi a \\
0 & i \left[\omega_p+\chi \left( c+c^*\right) \right]-\kappa & i \lambda & i \lambda  & i\chi a^* & i\chi a^*\\
-i \lambda & -i \lambda & -i \omega_{10} & 0 & 0 & 0\\
i \lambda & i \lambda & 0 & i \omega_{10} & 0 & 0\\
-i \chi a^* & -i \chi a & 0 & 0 & -i\omega_{20} & 0\\
i \chi a^* & i \chi a & 0 & 0 & 0 & i \omega_{20}
\end{pmatrix}.
\end{align}
To obtain the time-dependent Jacobian $\mathbf{J}_{\mathrm{LC}}(t)$, we substitute the mean-field LC solution for the dynamical variables $a(t)$, $b(t)$, and $c(t)$ into Eq.~\eqref{eq:j0}.
\end{widetext}


\section{Phase Diagram For the 2D Atom-Cavity Model}

A detailed discussion of the 2D atom-cavity model and the calculation of its mean-field dynamics can be found in Refs.~\cite{Cosme2019,Kongkhambut2024} (Refs.~[18,~19] in the main text). For our 2D simulations, we apply a linear ramp of the pump intensity $\epsilon$ for 5 ms and keep it constant at the desired value of $\epsilon$ for 25 ms.

\begin{figure*}[!ht]
\centering
\includegraphics[width=0.8\columnwidth]{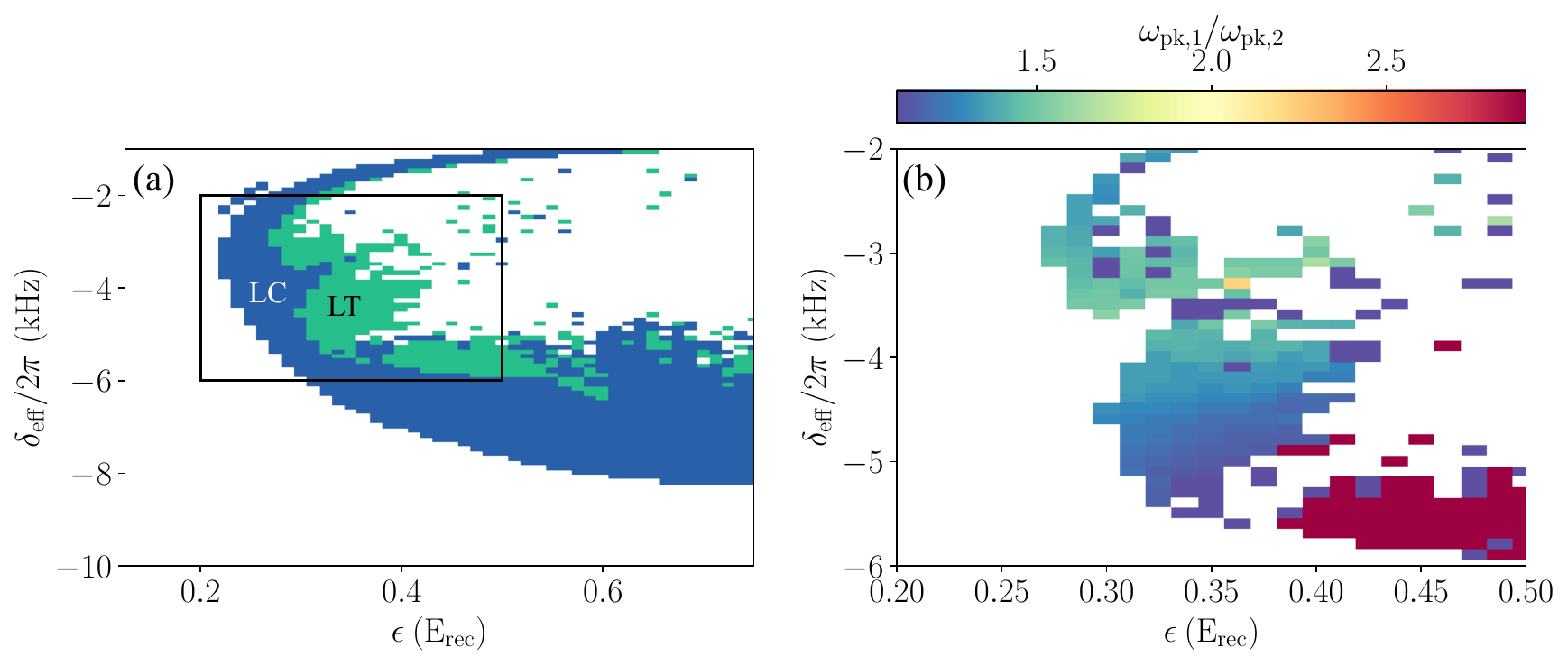}
\caption{(a) Theoretical phase diagram showing the LC and LT for a mean-field description of the 2D atom-cavity model. (b) Zoom-in of the area enclosed by the rectangle in (a) showing the ratio between the primary and secondary peak frequencies.} 
\label{sfig:3} 
\end{figure*}

To identify only the LC and limit torus (LT) solutions, we apply the following criteria. A solution is classified to be in the normal phase if the long-time averaged value of the intracavity photon number $|a|^2$ is below a threshold value, $|a|^2<10^{-3}$. 
The superradiant phase (SR) and chaotic dynamics (CH)  both possess a nontrivial Fourier spectrum of the photon dynamics $|{\alpha}|^2(\omega)=\mathcal{F}(|a|^2(t))$. To classify them, we use the relative crystalline fraction $\Xi$ defined as the value of the normalised spectrum $|\tilde{\alpha}|^2(\omega)=|{\alpha}|^2(\omega)/\sum_\omega |{\alpha}|^2(\omega)$ at the dominant or primary peak frequency $\omega_{\mathrm{pk},1}$, i.e., $\Xi = |\tilde{\alpha}|^2(\omega_{\mathrm{pk},1})=\mathrm{max}(|\tilde{\alpha}|^2)$. Removing the $\omega=0$ component, we identify a response as SR or CH if the maximum value of the relative crystalline fraction $\Xi$ is below a threshold value, $\Xi<10^{-4}$. For the range of values for $\delta_\mathrm{eff}$ and $\epsilon$ in Fig.~\ref{sfig:3}(a), a SR may exhibit transient oscillations, which decay for longer times, with oscillation frequencies between $100$ and $800$ Hz. To avoid misidentifying SR with transient oscillations as LC or LT, we also apply a lower bound for the dominant peak frequency $\omega_{\mathrm{pk},1}/2\pi \geq 1~\mathrm{kHz}$ for the LC and LT. A response is classified as a LC if $\Xi > 0.25$ and if the value of the normalised spectrum at the second or subdominant peak $\omega_{\mathrm{pk},2}$ is below a threshold value, $|\tilde{\alpha}|^2(\omega_{\mathrm{pk},2})<0.01$. On the other hand, for a LT, we check if the subdominant peak has $|\tilde{\alpha}|^2(\omega_{\mathrm{pk},2})\geq 0.01$. 

Applying the above criteria for LC and LT, we show in Fig.~\ref{sfig:3}(a) the phase diagram highlighting the LC and LT in the 2D atom-cavity model for the same set of parameters in the experiment. In Fig.~\ref{sfig:3}(b), we present the ratio between the primary and secondary peak frequencies $\omega_{\mathrm{pk},1}/\omega_{\mathrm{pk},2}$ for solutions identified as LT. We find that the characteristic of a quasiperiodic response, this ratio smoothly varies as $\omega_{\mathrm{pk},2}$ and $\omega_{\mathrm{pk},1}$ are incommensurate, in general, for a LT.

\section{Period-doubling Bifurcation}

\begin{figure*}[!ht]
\centering
\includegraphics[width=0.7\columnwidth]{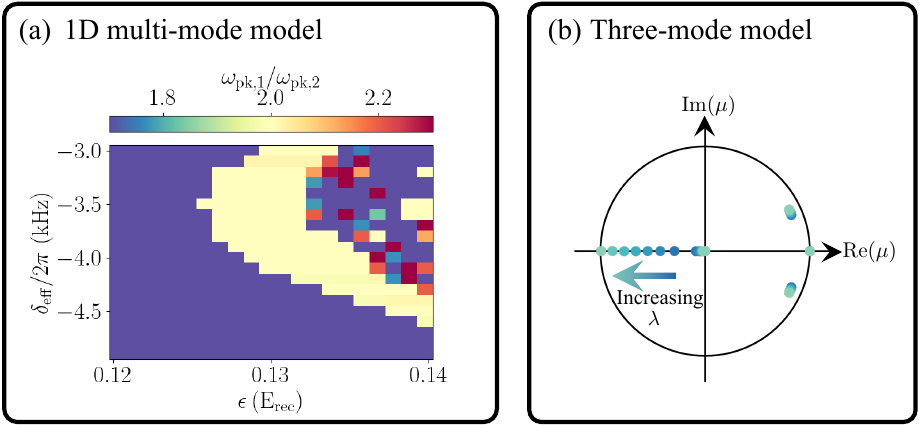}
\caption{Period doubling as seen from (a) the 1D atom-cavity model and (b) the minimal model of coupled bosonic modes. (a) The ratio between the primary and secondary peak frequencies $\omega_{\mathrm{pk},1}/\omega_{\mathrm{pk},2}$ in the 1D atom-cavity model. (b) Floquet multiplier for increasing light-matter coupling $\lambda$. Similar to the main text, we fix $\omega_p=\kappa$ in the minimal model.} 
\label{sfig:4} 
\end{figure*}

We also observe in our theoretical models a period-doubling bifurcation of the LC. This behaviour is found in the one-dimensional (1D) limit of the atom-cavity model, wherein only the dynamics along the cavity axis is considered. In Fig.~\ref{sfig:4}(a), we show the ratio $\omega_{\mathrm{pk},1}/\omega_{\mathrm{pk},2}$ for different combinations of $\delta_\mathrm{eff}$ and $\epsilon$. We find a large area in the parameter space, in which the ratio of the primary and secondary peak frequencies is $\omega_{\mathrm{pk},1}/\omega_{\mathrm{pk},2} = 2.0$, indicative of a period-doubling of the original LC. Moreover, $\omega_{\mathrm{pk},1}/\omega_{\mathrm{pk},2}$ does not smoothly vary with the parameters $\delta_\mathrm{eff}$ and $\epsilon$, unlike in the case of a Neimark-Sacker bifurcation in Fig.~\ref{sfig:3}(b). This points to a fundamentally different type of critical transition for the 1D case compared to the 2D or, in the case of the experiment, full three-dimensional system.

The 1D atom-cavity limit is modelled within the minimal bosonic system by simply adjusting the the transition frequency between the ground and first excited states of the atom from  $\omega_{10} = 2 \omega_\mathrm{rec}$ for the 2D case to $\omega_{10} = \omega_\mathrm{rec}$ for the 1D case, as only the momentum excitations along the cavity axis is included. Utilising Floquet stability analysis on the minimal model, we show in Fig.~\ref{sfig:4}(b) the behaviour of the Floquet multipliers and we observe that one of them crosses the unit circle as the LC loses its stability via a period-doubling bifurcation. The period-doubling bifurcation in this case is marked by a single real Floquet multiplier crossing the unit circle to $-1$ for increasing light-matter coupling strength as discussed in the main text.

\section{Importance of the third-level for the Neimark-Sacker Bifurcation}

As discussed in Ref.~\cite{Skulte2024}, the third atomic level associated with the $|0,\pm 2 \hbar k\rangle$ momentum modes can be adiabatically eliminated in the limit $\omega_{20} \gg \omega_{10}, \omega_p, \kappa$. This further simplifies the model to an effective nonlinear Dicke model, in which the photons experience a Kerr-like interaction \cite{Skulte2024}. Employing the Holstein-Primakoff transformation in the thermodynamic limit of this model, the equations of motion are \cite{Skulte2024}
\begin{align}\label{eq:eom2mode}
\frac{\mathrm{d} a}{\mathrm{d} t} &= -i \left[ \omega_p-i\kappa-2\frac{\chi^2}{\omega_{20}}|a|^2 \right] a-i\lambda \left(b+b^* \right),  \\ \nonumber
\frac{\mathrm{d} b}{\mathrm{d} t} &= -i \omega_{10} b -i \lambda  \left(a+a^* \right)~.
\end{align}

\begin{figure*}[!ht]
\centering
\includegraphics[width=0.7\columnwidth]{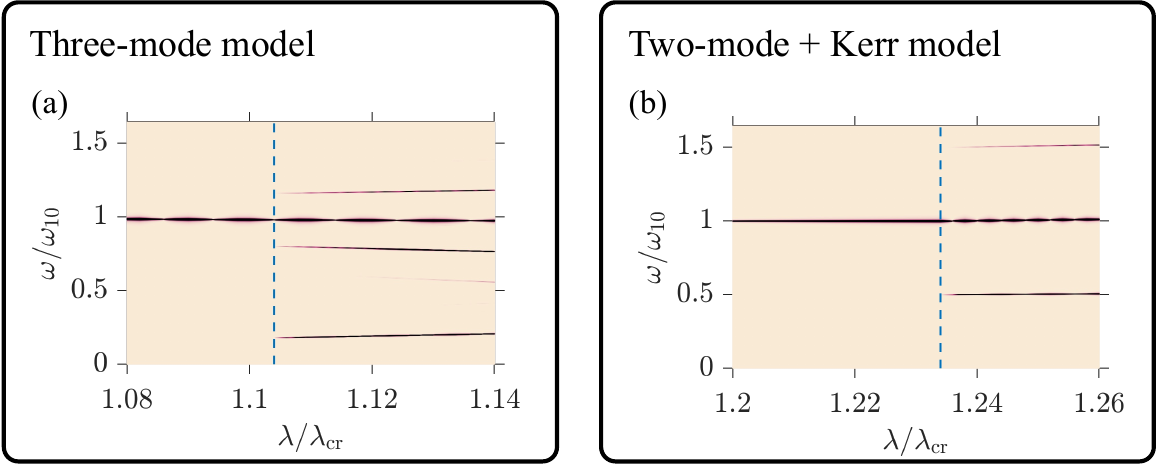}
\caption{Fourier spectra for (a) the three-mode model and the (b) the two-model model with a Kerr-like nonlinearity for the photons. The parameters are the same in both cases.} 
\label{sfig:5} 
\end{figure*}

The simplified two-mode model with a Kerr nonlinearity for the photons given by Eq.~\eqref{eq:eom2mode} correctly predicts a Hopf bifurcation for sufficiently large light-matter coupling, which amounts to an instability of the superradiant phase to form a limit cycle \cite{Skulte2024}. However, as shown in Fig.~\ref{sfig:5}, it misses on the prediction of a Neimark-Sacker bifurcation observed in the atom-cavity system. In Fig.~\ref{sfig:5}(a), the same Fourier spectrum shown in the main text is presented but now with the rescaled frequency axis to emphasize the appearance of incommensurate side peaks, which is a signature of a Neimark-Sacker bifurcation. In contrast, the spectra obtained from the two-mode model with a Kerr nonlinearity in Fig.~\ref{sfig:5}(b) displays a period-doubling bifurcation. Therefore, the presence of a third level in the atomic sector is crucial for the occurrence of a Neimark-Sacker bifurcation.

\section{Evidence for spontaneous symmetry breaking}
\begin{figure*}[!ht]
	\centering
	\includegraphics[width=0.7\columnwidth]{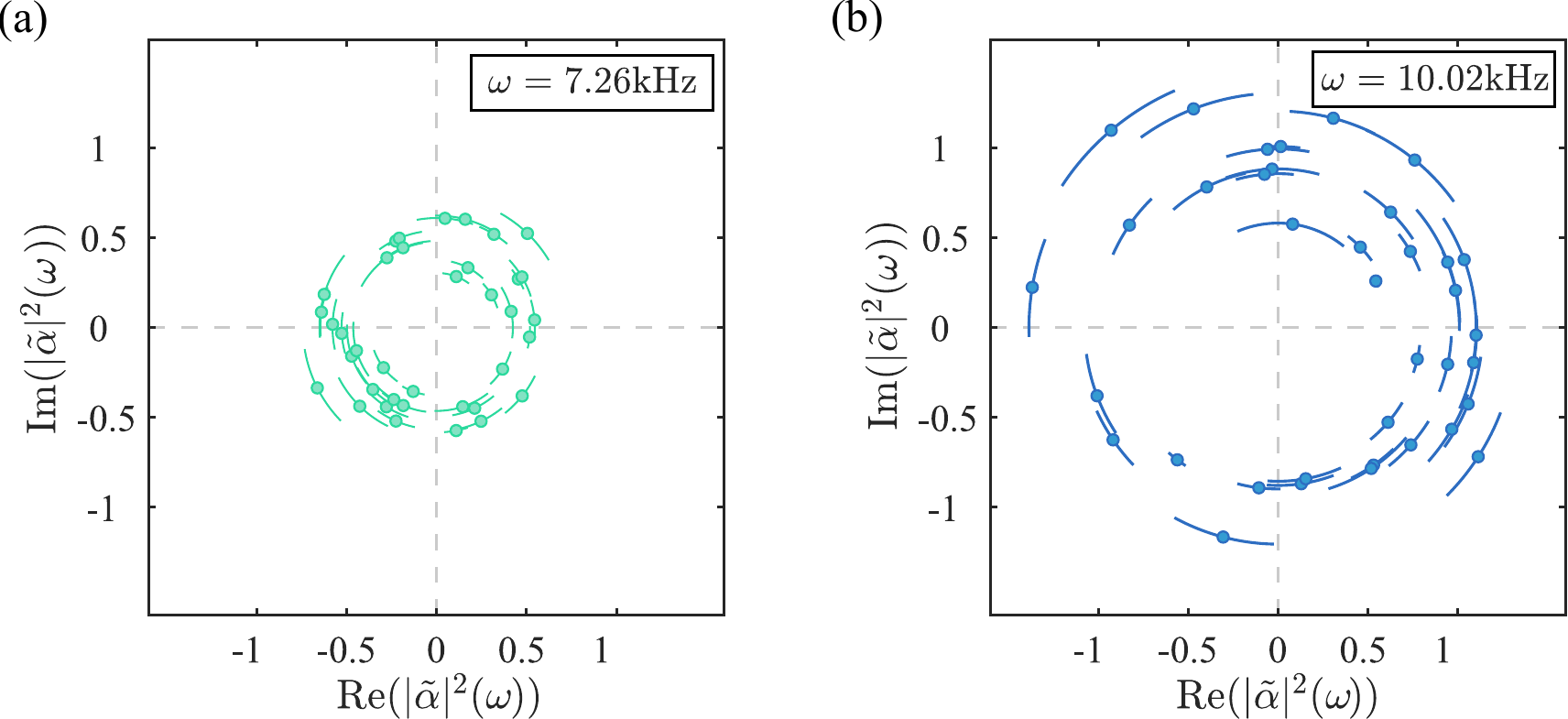}
	\caption{{Time phase distribution for the two main frequency components in the LT phase $\omega= 2\pi \times 7.26$~kHz (a) and $\omega= 2\pi \times 10.02$~kHz (b). For both frequencies the time phase difference between two realizations takes random values between $0$ and $2\pi$ as one would expect for spontaneous breaking of a continuous time translation symmetry.}}
	\label{sfig:6} 
\end{figure*}
{In the LT phase we observe that the intracavity photon number $|\alpha|^2$ oscillates with two main frequencies components, which is shown in Figs.~1(d) and (e) in the main text. To study these oscillations in the context of a quantum many-body system, which have time crystalline properties, we demonstrate strong evidence for spontaneous symmetry breaking with the measurements shown in Fig. 4. We prepare our experimental system in the LT phase, using an effective detuning $\delta_\mathrm{eff} = -2\pi \times 2.6$~kHz and a final pump strength  $\epsilon_\mathrm{f} = 1.15 E_\mathrm{rec}$, calculate the Fourier spectrum of the dynamics of $|\alpha|^2$, and plot the real and imaginary parts for the main frequency components as a polar plot. By repeating this experiment multiple times, we observe that the time phase, represented in the plot by the angle, takes random values between $0$ and $2\pi$, as one would expect for spontaneous breaking of a continuous time translation symmetry for both main frequency components in the spectrum $\omega= 2\pi \times 7.26$~kHz Fig.~\ref{sfig:6}(a) and $\omega= 2\pi \times 10.02$~kHz Fig.~\ref{sfig:6}(b), respectively.}

\section{Effects of particle loss and noise on the attractor}
\begin{figure*}[!ht]
	\centering
	\includegraphics[width=0.9\columnwidth]{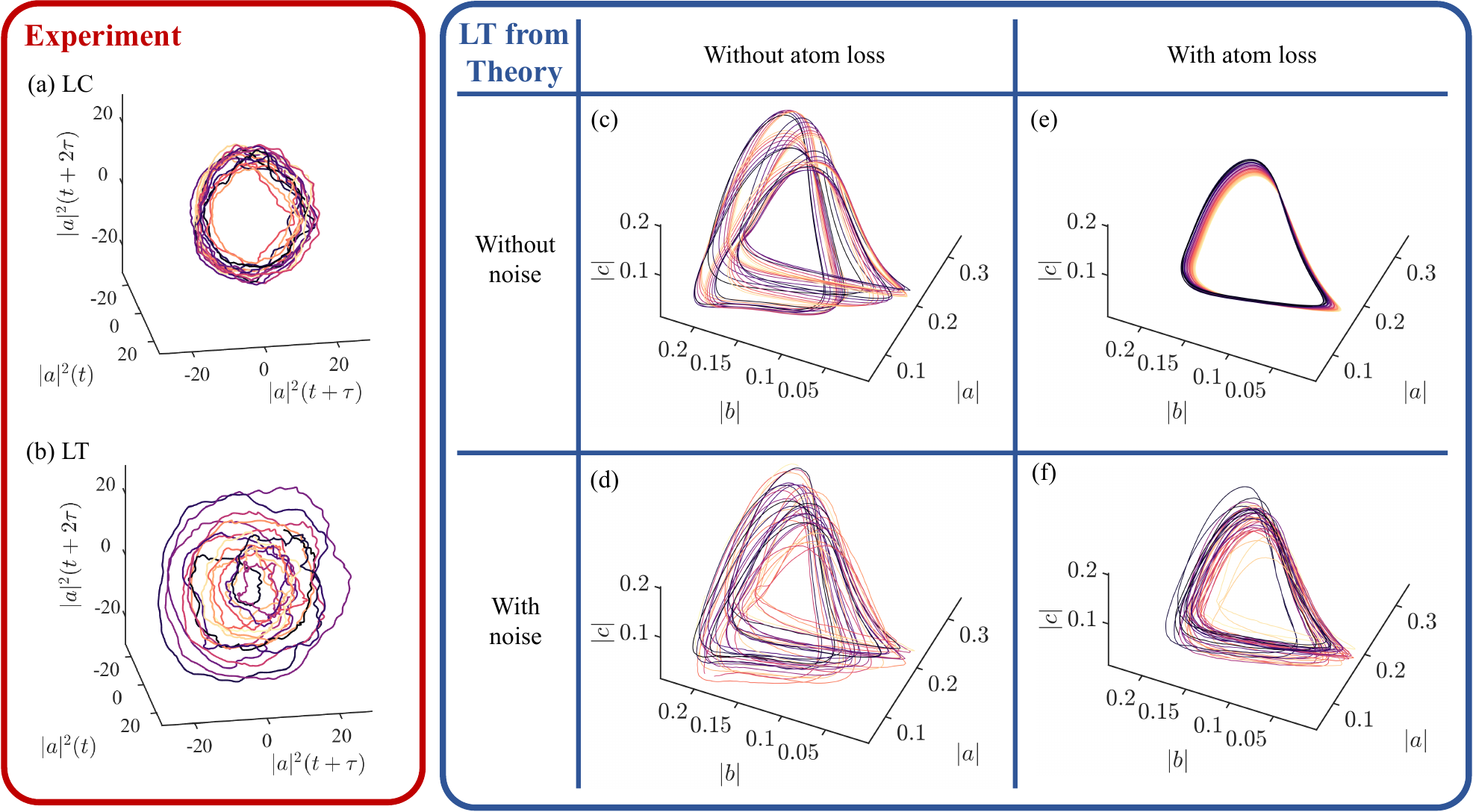}
	\caption{3D attractors from experimental data (left)and simulations (right). (a-b) the experimental LC and LT reconstructed attractors (Fig.~(3c) of the main text) with their time-axis information encoded in the color scale. From dark to light color: increasing time. (c-f) the simulated LT attractors with different conditions: with/without noise and atom loss.}
	\label{sfig:7} 
\end{figure*}
{In this section we show additional simulations supporting the experimental reconstructed attractors shown in Fig.~(3c) of the main text. In particular, we theoretically study the effects of particle loss and noise on the shape of the attractor and compare them with the attractor reconstructed from the experimental data. To allow for particle loss, we explicitly include the dependence of the equations of motion of the three-mode model on the particle number $N$ originating from the inherent parameters of the atom-cavity system. The relevant equations of motion are \cite{Tuquero2024}
\begin{align}
\frac{\partial a}{\partial t} &= -i\left[\omega_p - i \kappa + \chi N (c+c^*)\right] a - i \lambda(b+b^*) + \xi \\
\frac{\partial b}{\partial t} &= -i\omega_{10}b - i \lambda(a+a^*) \\
\frac{\partial c}{\partial t} &= -i\omega_{20}c - i \chi N a^*a
\end{align}
where the stochastic noise $\xi = \tilde{\xi}/\sqrt{N}$ due to photon loss follows $\langle \tilde{\xi}^*(t)\tilde{\xi}(t') \rangle = \kappa \delta(t-t')$. We model the atom-loss using a time-dependent particle number given by $N = N_0 \exp(t/\tau)$. In the following, we choose $N_0=40\times10^3$ and a value of $\tau$ such that the particle number decreases to $95\%$ of its initial value after 25 ms. A time-dependent $N$ not only changes the effective quantum noise in time and the strength of the nonlinearity as seen from above, but more importantly, it leads to a time-varying coupling strength and photon frequency, and thereby the effective detuning: $\lambda = \sqrt{\epsilon_{0} \omega_\mathrm{rec} N |U_0|}/2$ and $\delta_\mathrm{eff} = -\omega =\delta_c - NU_0/2$, where $\epsilon_{0}$ is the pump intensity assumed to be constant.

Fig.~\ref{sfig:7} shows both the experimental and theoretical attractors with their time-axis information encoded in the color scale, i.e., dark to light colour corresponds to increasing time. In the absence of quantum noise, the simulated torus with atom loss, Fig.~\ref{sfig:7}(c,e), becomes smaller in time as the system effectively gets pushed to the LC phase with decreasing particle number in time due to the effective reduction of $\delta_\mathrm{eff}$ and $\lambda$. This leads to the apparent spiralling of the trajectory towards the centre of the LT attractor, as can be seen in the theoretical results shown in Fig.~\ref{sfig:7}(e). The effect of the quantum noise manifests as fluctuations in the trajectory causing the attractor to occupy more volume in the phase space, see Fig.~\ref{sfig:7}(d). With both particle loss and quantum noise, Fig.~\ref{sfig:7}(f), the simulated attractor shrinks towards its centre and the trajectory occupies more phase space. Thus, we can attribute the apparent increase in the volume of the attractor in the experiment, Fig.~\ref{sfig:7}(b) to quantum noise and the spiralling of the trajectory at later times to atom losses.
	
\section{Decorrelator versus power spectrum}
We now discuss the challenge in obtaining the Lyapunov exponent and related quantities from the experimental data and why we opted to use information inferred from the power or Fourier spectrum, such as the amplitude of the subdominant peaks, as figures of merit for distinguishing the various dynamical phases in the system.

\begin{figure*}[!ht]
	\centering
	\includegraphics[width=0.8\columnwidth]{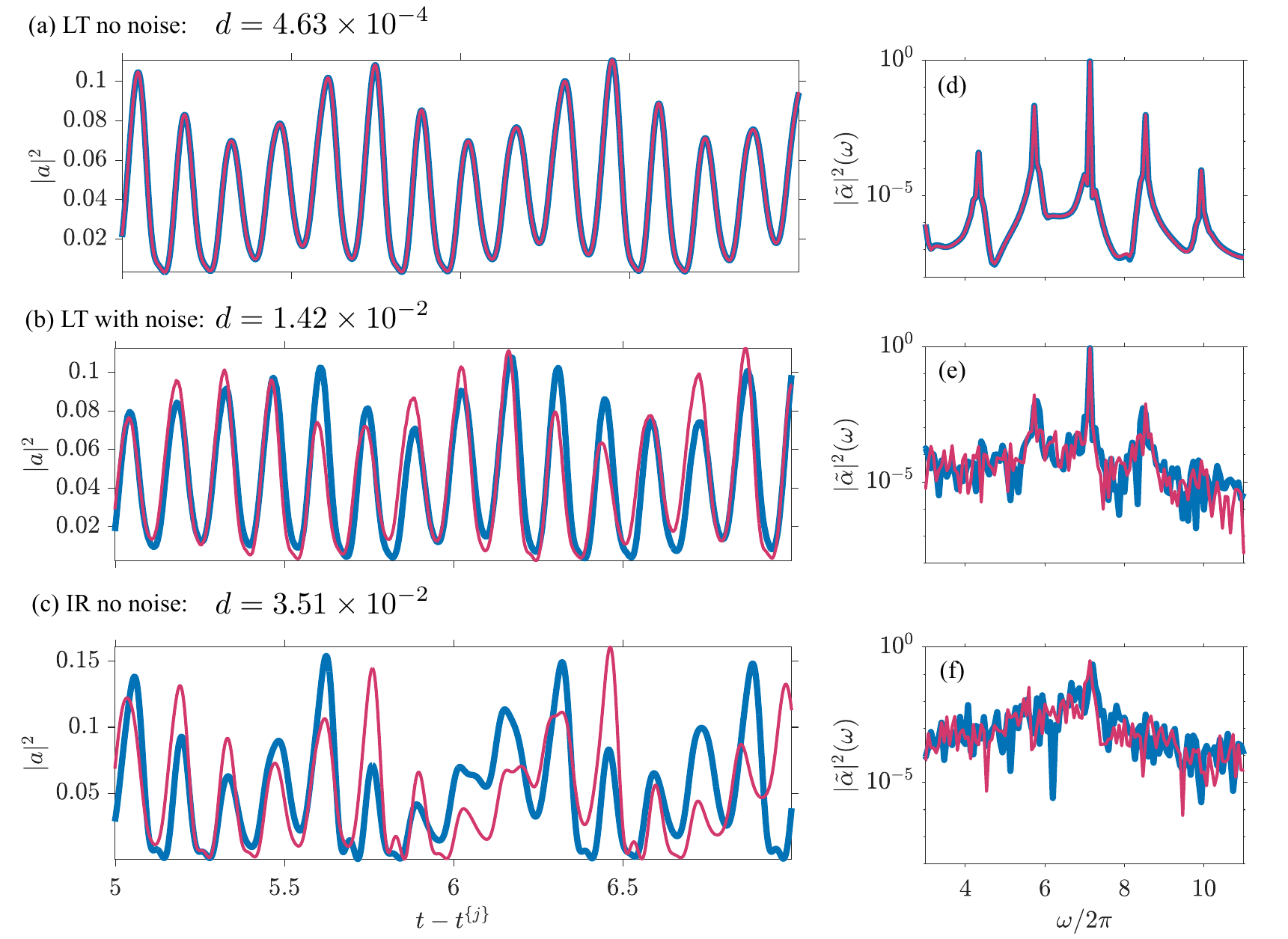}
	\caption{(a)-(c) Dynamics of the photon number starting from two slightly different initial states for (a) LT without noise, (b) LT with noise, and (c) IR without noise. The corresponding decorrelator values $d$ are also indicated. The curves are shifted in time relative to the time at which the first peak in the oscillating photon dynamics appears. (d)-(f) Corresponding power spectrum of the dynamics in (a)-(c).}
	\label{sfig:8} 
\end{figure*}

In lieu of the Lyapunov exponent, a related quantity that is easier to obtain is the decorrelator, which we define as
\begin{equation}\label{eq:deco}
d = \frac{1}{N_\mathrm{steps}}\sum_{i=1}^{N_\mathrm{steps}} \biggl| \left|a_1\left(t_i^{1}\right)\right|^2-\left|a_2\left(t_i^{2}\right)\right|^2 \biggr|.
\end{equation}
This definition of the decorrelator is related to the Lyapunov exponent at short times since $d \sim e^{\lambda t}$ \cite{Pizzi2021}. Here, $N_\mathrm{steps}$ is the number of recorded time steps in a chosen time window, which in the following simulations is $t\in [5,25]~\mathrm{ms}$. $a_1(t)$ and $a_2(t)$ are solutions for two different initial states differing by $1\times 10^{-5}$ for the initial values of $a$, $b$, and $c$ in the three-mode model. To obtain a small decorrelator for classifying regular dynamics, such as those for a LC or a LT, we made two trajectories line up by taking the time corresponding to the first peak of the photon number dynamics for each trajectory starting from 5 ms. We then use this as the initial times $t_i^{1}$ and $t_i^{2}$ when calculating the decorrelator in Eq.~\eqref{eq:deco}. In the absence of noise, the decorrelator value is small for a LT as depicted in Fig.~\ref{sfig:8}(a), which is consistent with the almost overlapping curves corresponding to the photon dynamics from two slightly different initial states. When quantum noise is included, the value of the decorrelator increases by two orders of magnitude as shown in Fig.~\ref{sfig:8}(b). This value of the decorrelator is already the same order of magnitude for an irregular phase in the absence of noise exemplified in Fig.~\ref{sfig:8}(c). This highlights the difficulty in obtaining reliable measures of irregularity like the Lyapunov exponent from realistic data containing both quantum and technical noise. In contrast, we find that the presence of side peaks in the power spectrum is more reliable for distinguishing LT and irregular dynamics as the noise simply broadens the characteristic frequency peaks in a LT as depicted in Figs.~\ref{sfig:8}(d)-(f).
	
\bibliography{references}